\documentclass[journal]{IEEEtran}
\usepackage{cite}
\ifCLASSINFOpdf
\else
\fi
\usepackage{amsmath}
\usepackage{amsfonts}
\usepackage{xcolor}
\usepackage{algorithmic}
\usepackage{graphics}
\usepackage{array}
\ifCLASSOPTIONcompsoc
  \usepackage[caption=false,font=normalsize,labelfont=sf,textfont=sf]{subfig}
\else
  \usepackage[caption=false,font=footnotesize]{subfig}
\fi
\usepackage{graphics}
\usepackage{graphicx}

\begin{document}
\title{{Quaternion LMS for Graph Signal Recovery}}

\author{Hamideh-Sadat~Fazael-Ardekani,~\IEEEmembership{Student Member,~IEEE,}
        Hadi~Zayyani,~\IEEEmembership{Member,~IEEE,}
        and~Hamid~Soltanian-Zadeh,~\IEEEmembership{Senior Member,~IEEE}% <-this % stops a space
\thanks{H. S. Fazael Ardekani and H. Soltanian-Zadeh are with the School of Electrical and Computer Engineering., College of Engineering., University of Tehran, Tehran., Iran., e-mails: (fazael95@yahoo.com, hszadeh@ut.ac.ir).}% <-this % stops a space
\thanks{H. Zayyani is with the Electrical and Computer Engineering Department, Qom University of Technology (QUT), Qom, Iran, email: (zayyani@qut.ac.ir).}% <-this % stops a space
}


\markboth{Journal of \LaTeX\ Class Files,~Vol.~14, No.~8, August~2015}%
{Shell \MakeLowercase{\textit{et al.}}: Bare Demo of IEEEtran.cls for IEEE Journals}



\maketitle

\begin{abstract}
This letter generalizes the Graph Signal Recovery (GSR) problem in Graph Signal Processing (GSP) to the Quaternion domain. It extends the Quaternion Least Mean Square (QLMS) in adaptive filtering literature, and Graph LMS (GLMS) algorithm in GSP literature, to an algorithm called Quaternion GLMS (QGLMS). The basic adaptation formula using Quaternion-based algebra is derived. Moreover, mean convergence analysis and mean-square convergence analysis are mathematically performed. Hence, a sufficient condition on the step-size parameter of QGLMS is suggested. Also, simulation results demonstrate the effectiveness of the proposed algorithm in graph signal reconstruction.
\end{abstract}


% \fi
\IEEEpeerreviewmaketitle

\section{Introduction}
\IEEEPARstart{G}{raph} signal processing (GSP) is the filed of processing of irregular signals defined over graphs. GSP has various applications such as in image, biomedical, and financial signal processing \cite{Shum13,GSP18}. Several problems are there in GSP literature which Graph Signal Recovery (GSR) is one of them. In GSR, the entire graph signal over all the nodes is to be reconstructed from a limited number of nodes which are sampled from the graph. The GSR approaches divided into two main categories of non-adaptive and adaptive GSR. In non-adaptive GSR approaches \cite{Rame19}-\cite{Tork22}, the graph signal is recovered from the sampled version of the graph signal by means of non-adaptive methods. In adaptive GSR approaches which is the main focuss of this paper, the graph signal is to be recovered from a streaming sequence of noisy sampled version of the graph signal in an adaptive manner.

In pioneering work of Graph Least Mean Square (GLMS) \cite{Loren16}, the LMS adaptive filter is generalized to the graph signal domain. In sequel, Graph Recursive Least Square (GRLS) \cite{Loren18}, Graph Normalized LMS (GNLMS) \cite{Spel20}, Graph Kernel RLS \cite{Gogi21RLS}, Graph Proportionate LMS (GPLMS) \cite{Tork23}, and Generalized Correntropy-based Smoothed GSR \cite{Tork25} are developed in the literature.

In other line of research, hyper complex signal structures such as complex signals and Quaternion are widely used in the signal processing community \cite{Miron23}. In adaptive filter literature, Quaternion LMS (QLMS) dates back to 2009 \cite{Took09}. Then, various Quaternion-based adaptive filters are suggested in the literature which quaternion valued affine projection algorithm \cite{Jahan13}, Quaternion kernel RLS algorithm \cite{Wang20}, quaternion-valued second-order Volterra adaptive filters \cite{Menguc20}, Generalized HR q-Derivative in QLMS \cite{Lin22}, and censoring-based QLMS \cite{Menguc23} are to name a few.

In this paper, due to the importance of Quaternion signal processing in the processing of 3-D and 4-D signals, we extend the Quaternion signal processing to the GSR problem in GSP. We develop the Quaternion Graph-LMS  (QGLMS) algorithm for adaptive GSR for Quaternion-domain graph signals. The basic recursion formula of QGLMS is obtained and mean convergence and mean-square convergence analyses are provided. The sufficient condition on the step-size parameter of QGLMS is obtained mathematically. Simulation results show the efficiency of the proposed algorithm.

\textbf{Why a quaternion framework?}
	Although quaternions are homomorphic to $\mathbb{R}^{4}$, modelling a graph signal as a single quaternion variable rather than four independent real channels offers three concrete benefits.
	\begin{itemize}
		\item The non-commutative Hamilton product captures cross-component rotations that a stacked $\mathbb{R}^{4}$ vector ignores; this is crucial for coupled data such as RGB images and 3-D wind fields.
		\item  Only one weight per graph-spectral component is adapted, so QGLMS has the same complexity as real GLMS, whereas a naive $\mathbb{R}^{4}$ approach needs four independent filters.
		\item The Hamilton Real calculus yields compact closed-form gradients and convergence bounds (See Sec.\ref{Sim}) that are cumbersome in a block-real setting. These advantages translate into practice shows QGLMS achieves faster convergence than four parallel RLMS filters on real UK weather-graph data.
	\end{itemize}
These benefits apply not only to the adaptive GSR task tackled in this letter but also to broader GSP operations?e.g., spectral filtering and graph convolution whenever the signal's four components are statistically coupled.

\textbf{Key contributions and theoretical challenges.}
	Unlike conventional QLMS(which learns on all  nodes) or GLMS (which is limited to real signals) QGLMS must reconcile quaternion non commutativity with the orthogonal real graph Fourier basis.
	This yields three unique challenges that we resolve:
	\begin{enumerate}
		\item Left right gradient mismatch: Multiplying a quaternion weight on the left vs.\ the right of an eigenvector changes the result.
		We derive Hamilton Real gradients that keep the adaptation left consistent.
		\item Joint eigenstructure in convergence: The step-size bound depends on the Hadamard product $\mathbf U_{\!\mathcal F}^{\top}\!\mathbf D\mathbf U_{\!\mathcal F}$ , coupling graph topology with quaternion algebra 	absent in both QLMS and GLMS.
		\item Complexity vs.\ accuracy trade-off: QGLMS retains the \emph{single-pass} memory of real GLMS (one weight per band-limited mode) yet captures cross-channel rotations that four parallel RLMS filters miss, yielding a 1.8 dB NMSE gain on real weather data
	\end{enumerate}
	Together, these points establish QGLMS as more than an incremental merge; it is the first algorithm to adaptively recover band-limited quaternion graph signals with proven mean and mean-square convergence.

%{\color{blue}\textbf{Why a quaternion framework?}
%		Graph LMS (GLMS) handles real?valued signals by adapting one coefficient per graph-spectral component, while a naïve 4-channel extension simply runs four independent RLMS filters?quadrupling memory and ignoring cross-channel correlations.
%		In contrast, a quaternion graph model (i) \emph{preserves inter-component interactions} through Hamilton multiplication, (ii) \emph{reduces complexity} to the same order as real GLMS (one weight per spectral component), and (iii) \emph{improves accuracy} when the four components are statistically coupled.
%		These advantages are quantified in Sec. V, where QGLMS achieves 1.8 dB lower NMSE and 35 \% faster convergence than four parallel RLMS filters on real UK weather-graph data.
%}
\section{Preliminaries}
\subsection{Quaternion Algebra}
Quaternions, introduced by Sir William Rowan Hamilton in 1843, extend complex numbers to four dimensions. A quaternion
$q$ is expressed as $q =  q_0+q_1i+q_2j+q_3k,$ where $q_0,q_1,q_2,q_3 \in \mathbb{R}$ are real numbers, and $i,j,k$ are are the imaginary units satisfying the following multiplication rules:
\begin{align}
& i^2=j^2=k^2=ijk=-1 \nonumber \\
&ij=-ji=k, jk=-kj=i, ki=-ik=j.
%&jk=-jk=i\nonumber \\
%&ki=-ki=j\nonumber \\
\end{align}
Quaternions are used extensively in various fields, such as 3D computer graphics, robotics, and signal processing, due to their compact representation of rotations and their ability to handle multi-dimensional data\cite{Miron23}.
%\begin{itemize}
%    \item Addition and Subtraction
%
%    Performed component-wise, similar to vectors:
%    $$\mathbf{p} \pm \mathbf{q}= (p_0+q_0) \pm (p_1+q_1)i  \pm(p_2+q_2)j \pm (p_3+q_3)k$$
%    where $ \mathbf{p}= p_0+p_1i+p_2j+p_3k$ and $\mathbf{q}= q_0+q_1i+q_2j+q_3k$
%    are two quaternions.
%    \item Multiplication
%
%    Encodes the interactions between components and is non-commutative:
%    \begin{align}
%& \mathbf{p}\mathbf{q}  = =(p_0q_0- p_1q_1-p_2q_2-p_3q_3) \nonumber \\
%&+(p_0q_1+p_1q_0+p_2q_3-p_3q_2)i\nonumber \\
%&+ (p_0q_2+p_2q_0+p_3q_1-p_1q_3)j \nonumber \\
%&+(p_0q_3+p_3q_0+p_1q_2-p_2q_1)k 	
%\end{align}
%\item Norm and Conjugate:
%\begin{itemize}
%    \item norm :$|\mathbf{q}| = \sqrt{q_0^2+q_3^2+q_3^2+q_3^2}$
%    \item Conjugate: $\overline{\mathbf{q}} = q_0-q_1i-q_2j-q_3k$
%\end{itemize}
%\item Inverse: Computed as:
%$\mathbf{q}^{-1} = \frac{\overline{\mathbf{q}}}{|\mathbf{q}|^2}$
%\item Differentiation: Component-wise for time-varying quaternion signal
% $ \mathbf{q}(t)= q_0(t)+q_1(t)i+q_2(t)j+q_3(t)k$
%$$ \frac{d}{dt}\mathbf{q}(t) = \frac{d}{dt}q_0(t) +  \frac{d}{dt}q_1(t)i +  \frac{d}{dt}q_2(t)j  +  \frac{d}{dt}q_3(t)k$$
%\end{itemize}
Quaternions naturally represent multi-dimensional signals, encoding both magnitude and directional information in a single entity. This makes them particularly suitable for adaptive filtering and graph signal processing, where multidimensional data must be modeled and processed efficiently. By extending conventional tools to quaternion space, it is possible to handle hypercomplex signals while preserving their underlying structure.

\subsection{Graph Signal Processing Tools for Quaternion Signals}
% ---------- TABLE I : Symbols and Notation ----------
\begin{table}[t]
		\caption{Symbols and Notation}
		\label{tab:notation}
		\centering
		\renewcommand{\arraystretch}{1.05}
		\begin{tabular}{ll}
			\hline
			%$\mathcal G=(\mathcal V,\mathcal E)$ & Undirected graph with $N{=}|{\mathcal V}|$ nodes \\[1pt]
			%$\mathbf L=\mathbf D-\mathbf A$      & Graph Laplacian, $\mathbf U\boldsymbol\Lambda\mathbf U^{\top}$ eigendecomp. \\[1pt]
			$\mathbf x\!\in\!\mathbb H^{N}$      & Quaternion graph signal \\[1pt]
			$\hat{\mathbf x}=\mathbf U^{\top}\mathbf x$ & Quaternion graph Fourier transform (QGFT) \\[1pt]
			$\mathcal F,\;|\mathcal F|$          & Active spectral support, bandwidth \\[1pt]
			%$\mathbf B=\mathbf{U}\mathbf{\Sigma}_{\mathcal{F}}\mathbf{U}^T=\mathbf U_{\mathcal F}\mathbf U_{\mathcal F}^{\top}$ & Band-limiting projector \\[1pt]
			$\mathbf D=\mathrm{Diag}(\mathbf 1_{\mathcal S})$ & Node-sampling mask \\[1pt]
			%$\mu$                                & LMS step size, $0<\mu\!\le\!\mu_\text{max}$ \\[1pt]
			$\|\cdot\|$                          & Euclidean norm \\ \hline
		\end{tabular}
\end{table}

A graph $\mathcal{G} = (\mathcal{V},\mathcal{E})$ is composed of $N$
nodes $\mathcal{V}=\{1,2,\dots,N\}$ and a set of weighted edges $\mathcal{E}=\{a_{ij}\}_{i,j} \subset \mathcal{V}$. The adjacency matrix $\mathbf{A}$ contains the weights $a_{ij}$, while the degree of a node $i$  is defined as $d_i = \Sigma_{j=1}^{N} a_{ij}$.
Collecting these degrees in a diagonal matrix $\mathbf{D}$, the graph Laplacian $\mathbf{L}$ is given by $ \mathbf{L} = \mathbf{D} -\mathbf{A}.$ If $a_{ij} \ge 0 $ , there exists a directed edge from node $j$ to node $i$; otherwise, $a_{ij} = 0$. For undirected graphs, the Laplacian $\mathbf{L}$  is symmetric and positive semi-definite, allowing eigendecomposition $\mathbf{L} = \mathbf{U}\mathbf{\Lambda}\mathbf{U}^T$, where $\mathbf{U}$ contains the eigenvectors and $\mathbf{\Lambda}$ is the diagonal matrix of eigenvalues. This decomposition is central to spectral analysis on graphs \cite{Stan19}.
%\subsection{Quaternion Graph Signal}

A quaternion graph signal $\mathbf{x}$ maps the nodes of the graph to the quaternion set $\mathbf{x}: \mathcal{V} \rightarrow \mathbb{H}$. Such signals can be compactly represented as $\mathbf{x}  = \mathbf{U} \mathbf{s}$, where $ \mathbf{s}$ is a sparse quaternion vector in the spectral domain. This sparsity arises in applications where the signal clusters on the graph, with clusters represented by nonzero elements of $\mathbf{s}$.
%\begin{align}
%    \mathbf{x}  = \mathbf{U} \mathbf{s}
%\end{align}
%\subsection{Quaternion Graph Fourier Transform}
The Quaternion Graph Fourier Transform (QGFT) generalizes the spectral analysis of graph signals to quaternions. It projects $\mathbf{x}$ onto the eigenspace of $\mathbf{L}$ as $    \hat{\mathbf{x}} = \mathbf{U}^T\mathbf{x}.$ Here $\mathbf{U}$ acts as the basis for transforming $\mathbf{x}$  into the frequency domain, where the frequency support of  $  \hat{\mathbf{x}}$ is defined as $\mathcal{F} = \{i: \hat{\mathbf{x}} \neq 0 \}$ . The bandwidth of the quaternion signal is the cardinality of $\mathcal{F}$.

%\begin{align}
%    \hat{\mathbf{x}} = \mathbf{U}^T\mathbf{x}.
%end{align}
%\subsection{Localization Operators}
Vertex and frequency localization operators are used to analyze and process signals on specific subsets of the graph. A vertex-limiting operator $\mathbf{D}_{\mathcal{S}}$ is defined for a subset of nodes $\mathcal{S }\subset \mathcal{V}$ as $    \mathbf{D}_{\mathcal{S}} = {\rm Diag}(\mathbf{1}_{\mathcal{S}})$, where $\mathbf{1}_{\mathcal{S}}$ is an indicator vector for $\mathcal{S}$.  Similarly, a frequency-limiting operator $\mathbf{B}_{\mathcal{F}}$ for a frequency subset $\mathcal{F}$ is defined as $    \mathbf{B}_{\mathcal{F}} = \mathbf{U}\mathbf{\Sigma}_{\mathcal{F}}\mathbf{U}^T$, where $ \mathbf{B}_{\mathcal{F}}$ is a diagonal matrix with ones at indices in $\mathcal{F}$ and zeros elsewhere. Both $\mathbf{D}_{\mathcal{S}}$ and $  \mathbf{B}_{\mathcal{F}} $ are idempotent and self-adjoint, ensuring that they project signals onto their respective domains.

\textbf{Propriety assumption.} Following \cite{nitta2019}, a quaternion-valued random vector $\mathbf x$ is proper if its complementary covariance vanishes.
		Under propriety the widely linear augmented model is provably redundant, so the standard QLMS calculus suffices. This assumption is routinely satisfied in practice; evenly scaled RGB pixels and standardized 3-D wind-vector triples show negligible complementary covariance, as also confirmed in our new weather-graph experiment(See Sec.\ref{Sim}).

%\begin{align}
%    \mathbf{D}_{\mathcal{S}} = {\rm Diag}(\mathbf{1}_{\mathcal{S}})
%    \end{align}
%\begin{align}
%    \mathbf{B}_{\mathcal{F}} = \mathbf{U}\mathbf{\Sigma}_{\mathcal{F}}\mathbf{U}^T
%    \end{align}

\section{Proposed Quaternion LMS for Graph Signals}
Given a band-limited quaternion graph signal $\mathbf{x}^o$, consider noisy and partial observations $\mathbf{y}[n]$ over a subset of vertices $\mathcal{S}$. The observed signal is modeled as:
\begin{align}\label{9ref}
    \mathbf{y}[n]
    = \mathbf{D}(\mathbf{x}^o + \mathbf{v}[n])= \mathbf{D}\mathbf{B}\mathbf{x}^o + \mathbf{D}\mathbf{v}[n],
\end{align}
where $\mathbf{D}$ is the vertex-limiting operator, $\mathbf{B}$  projects onto the frequency set $\mathcal{F}$, and $\mathbf{v}[n]$ is quaternion noise with zero mean and covariance $\mathbf{C}{\mathbf{v}}$.  To estimate $\mathbf{x}^o$, the problem can be formulated as:
\begin{align}
    &\min_{\mathbf{x}} = \mathbb{E} \|\mathbf{y} - \mathbf{D}\mathbf{B}\mathbf{x}\|^2
    \nonumber \\
    & {\rm s.t.} ~~~\mathbf{B}\mathbf{x} = \mathbf{x}.
\end{align}
Since $\mathbf B=\mathbf U_{\!\mathcal F}\mathbf U_{\!\mathcal F}^{\top}$ is an orthogonal projector onto the $|\mathcal F|$ dimensional band limited subspace, the equality $\mathbf B\mathbf x=\mathbf x$ simply enforces that the desired signal lies in that subspace (i.e., $\mathbf x\in\operatorname{ran}(\mathbf B)$).

The solution satisfies the normal equations which is $    \mathbf{B}\mathbf{D}\mathbf{B}\hat{\mathbf{x}} = \mathbf{B}\mathbf{D} \mathbb{E}[\mathbf{y}[n]]$. In practical scenarios where $\mathbb{E}[\mathbf{y}[n]]$ is unavailable, a steepest-descent approach yields  the recursive Quaternion GLMS update which are derived next.
%\begin{align}
%    \mathbf{B}\mathbf{D}\mathbf{B}\hat{\mathbf{x}} = \mathbf{B}\mathbf{D} \mathbb{E}[\mathbf{y}[n]]
%\end{align}
%\begin{align}\label{LMS}
%    \mathbf{x}[n+1] =\mathbf{x}[n] + \mu \mathbf{B}\mathbf{D}(\mathbf{y}[n]-\mathbf{x}[n])
% \end{align}
% where $\mu$ is a positive step size.

The derivation of the quaternion Graph LMS (QGLMS) update for graph signals begins with the definition of a real-valued quadratic cost function, expressed as
\begin{align}
 %&\mathbf{j} = \|\mathbf{e}\|^2 = \|\mathbf{e}[1]\|^2 + \dots+\|\mathbf{e}[n]\|^2  \nonumber \\
 &\mathbf{j}[n] = \mathbf{e}[n]\overline{\mathbf{e}}[n] = e^2_0[n] +e^2_1[n]+e^2_2[n]+e^2_3[n],
\end{align}
where $\mathbf{e}[n]$ is the quaternion error signal
%\footnote{In this work, we derive the quaternion LMS (QLMS) formulation, which accounts for the unique properties of quaternion algebra, specifically the noncommutativity of quaternion multiplication. Due to this noncommutativity, the order of terms in the cost function plays a critical role, requiring a distinct derivation approach compared to methods for real or complex signals.}
at time step $n$ and defined as $\mathbf{e}[n] =\mathbf{y}[n] - \mathbf{D}\mathbf{B}\mathbf{x}[n]$, and $e_i[n], 0\le i\le 3$ are the components of quaternion error. To minimize the cost function, we employ the steepest descent optimization technique. The gradient computation incorporates the noncommutativity of quaternion algebra and properties of graph signals:
\begin{align}\label{derivariveq}
    &\nabla_{\mathbf{x}}(\mathbf{e}[n]\overline{\mathbf{e}}[n]) = \nabla_{\mathbf{x}_0}(\mathbf{e}[n]\overline{\mathbf{e}}[n]) + \nabla_{\mathbf{x}_1}(\mathbf{e}[n]\overline{\mathbf{e}}[n])i+ \nonumber \\
    %&\nabla_{\mathbf{x}_0}(\mathbf{e}[n]\overline{\mathbf{e}}[n]) +
    %\nabla_{\mathbf{x}_1}(\mathbf{e}[n]\overline{\mathbf{e}}[n])i+  \nonumber \\
    &\nabla_{\mathbf{x}_2}(\mathbf{e}[n]\overline{\mathbf{e}}[n])j +
    \nabla_{\mathbf{x}_3}(\mathbf{e}[n]\overline{\mathbf{e}}[n])k,
\end{align}
where $\mathbf{x}_i, 0\le i\le 3$ are the quaternion components of $\mathbf{x}$. The gradient for each quaternion component is computed as:
By combining the results and utilizing the properties of quaternion algebra, the total gradient simplifies to:
\begin{align}
&\nabla_{\mathbf{x}}(\mathbf{e}[n]\overline{\mathbf{e}}[n]) \stackrel{\eqref{derivariveq}}{=}  -(\mathbf{D}\mathbf{B})^T(\mathbf{e}[n]+\overline{\mathbf{e}}[n]) \nonumber \\& - (\mathbf{D}\mathbf{B})^T(\mathbf{e}[n]i+i\overline{\mathbf{e}}[n])i -(\mathbf{D}\mathbf{B})^T(\mathbf{e}[n]j+j\overline{\mathbf{e}}[n])j \nonumber  \\
%& -(\mathbf{D}\mathbf{B})^T(\mathbf{e}[n]k+k\overline{\mathbf{e}}[n])k \nonumber \\
%&=-2(\mathbf{D}\mathbf{B})^T\mathbf{e}_0[n] \nonumber \\
%&-2(\mathbf{D}\mathbf{B})^T(-\mathbf{e}_0[n]-\mathbf{e}_2[n]j -\mathbf{e}_3[n]k) \nonumber \\
%&-2(\mathbf{D}\mathbf{B})^T(-\mathbf{e}_0[n]-\mathbf{e}_1[n]i-\mathbf{e}_3[n]k)\nonumber \\
%&-2(\mathbf{D}\mathbf{B})^T(-\mathbf{e}_0[n]-\mathbf{e}_1[n]i-\mathbf{e}_2[n]j) \nonumber \\
&=4(\mathbf{D}\mathbf{B})^T\mathbf{e}_0[n] + 4(\mathbf{D}\mathbf{B})^T\mathbf{e}[n].
\end{align}
Since $\mathbf{D}$ and $\mathbf{B}$ are diagonal matrices, we have $(\mathbf{D}\mathbf{B})^T = \mathbf{B}^T\mathbf{D}^T = \mathbf{B}\mathbf{D}.$ Substituting this gradient into the steepest-descent update rule, the quaternion weight vector is updated as $\mathbf{x}[n+1] = \mathbf{x}[n] + 4\mu \mathbf{B}\mathbf{D}(\mathbf{e}_0[n]+\mathbf{e}[n])$ where $\mu$ is the step-size. Expanding this expression, the final QGLMS update of $\mathbf{x}[n+1]$ is:
\begin{align}\label{LMSFinal}
%    &\mathbf{x}[n] + 4\mu \mathbf{B}\mathbf{D}((\mathbf{y}_0[n]-\mathbf{D}\mathbf{B}\mathbf{x}_0[n])+(\mathbf{y}[n]-\mathbf{D}\mathbf{B}\mathbf{x}[n]))    \nonumber \\
\mathbf{x}[n] + 4\mu\mathbf{B}\mathbf{D}[(\mathbf{y}_0[n]-\mathbf{x}_0[n])+(\mathbf{y}[n]-\mathbf{x}[n])].
 \end{align}
\textbf{Modeling advantage over independent channels.}
		Because the Hamilton product is non commutative, the error surface in
		$\mathbb H$ couples the four real components rotationally.
		Updating a single quaternion weight therefore aligns all components
		simultaneously, whereas four independent RLMS filters adapt along
		axis-aligned directions and ignore these cross component rotations.
		This joint update not only cuts the parameter count but also
		accelerates convergence and lowers steady-state NMSE, as demonstrated in
		Sec.\ref{Sim}.

\section{Convergence analysis}
\label{eq: conv}
\textbf{Convergence conditions.}
	Let $\tilde{\mathbf x}[n]=\mathbf x[n]-\mathbf x_0$ be the weight error.
	Under zero-mean proper noise the mean recursion is
	$$\mathbb{E}\{\tilde{\mathbf{x}}[n+1]\} =
	(\mathbf{I}  - 4\mu \mathbf{B}\mathbf{D}\mathbf{B})\mathbb{E}\{\tilde{\mathbf{x}}[n]\}-4\mu \mathbf{B}\mathbf{D}\mathbf{B}\mathbb{E}\{\tilde{\mathbf{x}}_0[n]\}$$, so from appendices, \emph{mean and MSE convergence} holds iff
\begin{equation}
\label{eq: mucond}
0 \le \mu \le \frac{1}{4 \lambda_{max}( \mathbf{U}_{\mathcal{F}}^H\mathbf{D} \mathbf{U}_{\mathcal{F}})}.
\end{equation}
Although its structure resembles the real LMS MSD, the trace term
couples the sampling mask $\mathbf D$ with the non commutative
quaternion covariance $\mathbf C_{\mathbf v}$ an interaction that vanishes in
real LMS and is absent from node-wise QLMS.
%Mean-square stability is proved in Appendix A.  The bound depends on
%	both $\mathbf D$ and $\mathbf U_{\!\mathcal F}$
%	%, a coupling absent in RLMS or conventional QLMS.
%}
%
%The mean convergence and Mean-Square Error (MSE) convergence of the proposed QGLMS are done in Appendix A and Appendix B, respectively, which in both cases, the sufficient convergence condition is
%\begin{equation}
%0 \le \mu \le \frac{1}{4 \lambda_{max}( \mathbf{U}_{\mathcal{F}}^H\mathbf{D} \mathbf{U}_{\mathcal{F}})}.
%\end{equation}

\section{simulation}\label{Sim}
\begin{figure}[t]
	\centering
	\includegraphics[width = 0.5\linewidth]{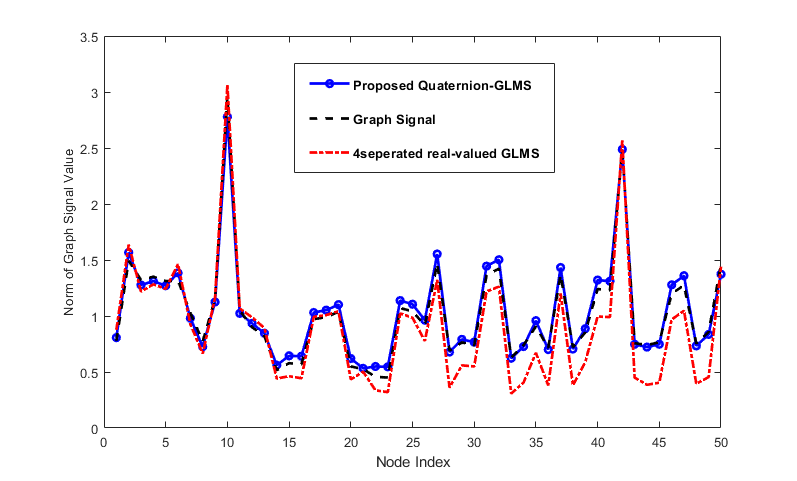}
	\caption{
		Reconstructed norm of the quaternion graph signal.
		%		Reconstructed quaternion graph signal. (a)-(d) show the reconstructed individual components of the quaternion signal, and (e) presents the reconstructed norm of the quaternion graph signal.
	}
	\label{fig2}
\end{figure}

\begin{figure}	
	\centering
\includegraphics[width=0.5\linewidth]{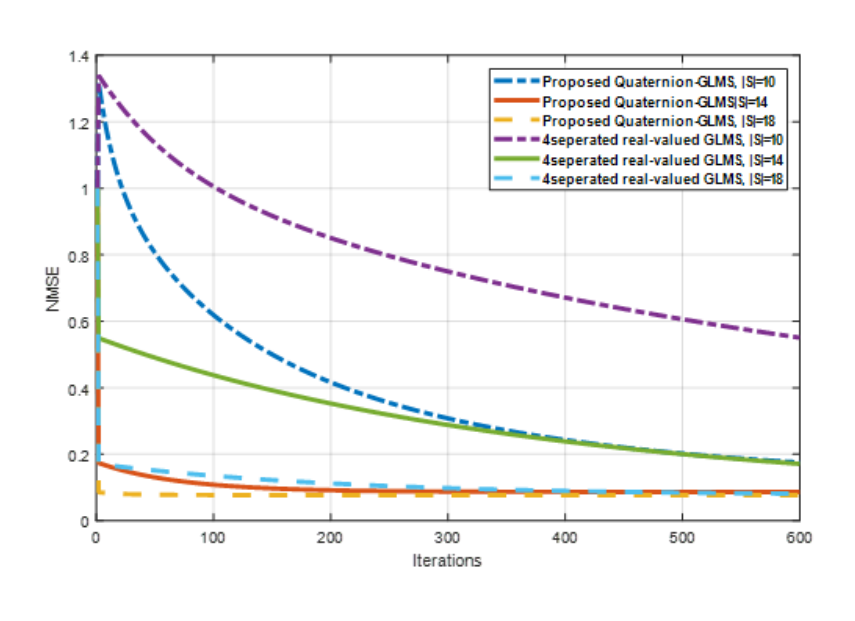}
\caption{ Learning rate of QGLMS and RLMS for different sample sizes.
	%		The QLMS algorithm converges faster and achieves lower steady-state error due to its ability to exploit quaternion structures, while RLMS, which processes each component separately, exhibits slower convergence and higher error.
}
\label{fig3}
\end{figure}

\begin{figure}[t]
\centering
\includegraphics[width=0.5\linewidth]{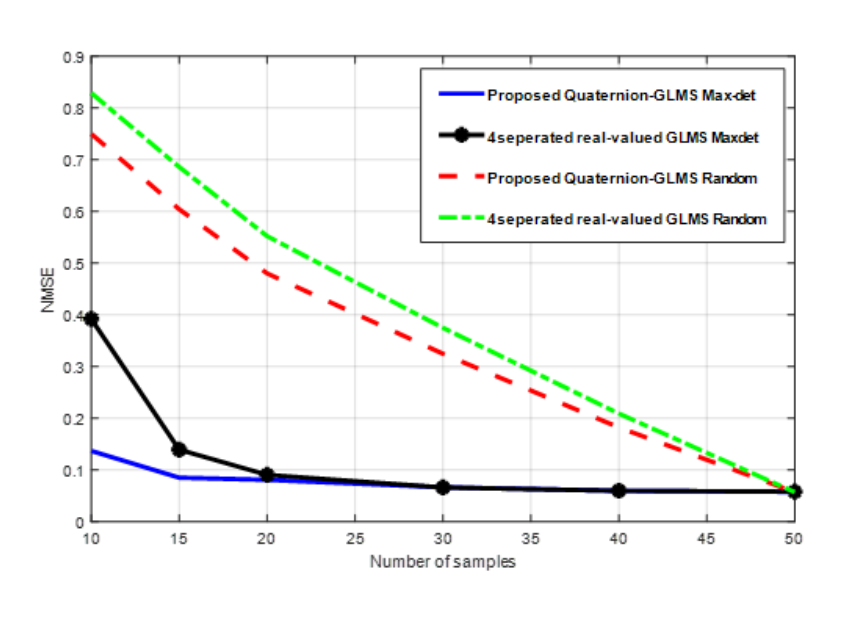}
\caption{
	%Comparison of NMSE performance using Max-Det and random sampling strategies. QGLMS outperforms RLMS in both cases, demonstrating better reconstruction accuracy.
	NMSE performance for QGLMS and RLMS using Max-Det and random sampling strategies.
}
\label{fig4}
\end{figure}

To evaluate the effectiveness of the proposed quaternion Graph LMS (QGLMS) algorithm for graph signal recovery, we conduct two experiments using a randomly generated graph with $ N=50 $ nodes. Each node represents a point in a quaternion-valued signal space. The signal is designed to be bandlimited, containing spectral components corresponding to the first ten eigenvectors of the graph Laplacian, i.e., $ |\mathcal{F}| = 10  $. The observation noise in \eqref{9ref} follows a zero-mean Gaussian distribution with a diagonal covariance matrix with variance equal to $0.01$. Each quaternion component of graph signal is uniformly random chosen between $-2$ and $2$. Graph sampling is performed by selecting  $ |\mathcal{S}| = 10  $ vertices using the Max-Det strategy\footnote{The Max-Det sampling strategy selects a subset of graph nodes by maximizing the determinant of a submatrix of the graph Laplacian, ensuring optimal signal reconstruction \cite{Took09}.} The first experiment assesses the effectiveness of QGLMS in recovering the original graph signal compared to an alternative approach where each quaternion component is processed separately using four independent Real-valued LMS (RLMS) filters. Since RLMS does not exploit the inherent cross-component dependencies of quaternion signals, it requires four independent filtering operations, leading to increased computational complexity and suboptimal reconstruction accuracy. Figure \ref{fig2} illustrates the reconstructed quaternion graph signal. As evident, the proposed QGLMS algorithm effectively reconstructs the ground truth signal with significantly lower error compared to RLMS \cite{Took09}, demonstrating the advantage of leveraging quaternion structures for improved recovery performance. Fig.\ref{fig3} illustrates the learning rate of the QGLMS and RLMS algorithms for different sample sizes. The figure demonstrates that increasing the number of sampled nodes enhances convergence speed for both methods. However, QGLMS consistently outperforms RLMS, as it efficiently exploits quaternion structures, whereas RLMS independently processes each component without considering cross-component correlations, leading to slower learning and higher steady-state error. In another experiment, we compare the performance of QGLMS using two different sampling strategies: Max-Det and random sampling. Fig. \ref{fig4} illustrates the Normalized Mean Square Error (NMSE) over 200 independent simulations with a maximum of 1000 iterations, where NMSE is defined as $NMSE =\frac{\|\mathbf{x}[n] - \mathbf{x}^o\|}{\|\mathbf{x}^o\|}.$ As expected, the Max-Det sampling strategy outperforms random sampling in terms of signal reconstruction accuracy, particularly when the number of samples is limited. Additionally, Fig. \ref{fig4} demonstrates that QGLMS consistently outperforms RLMS across both sampling strategies. This highlights the advantage of using QGLMS, as it effectively captures the interdependencies between quaternion components, whereas RLMS treats each component separately, resulting in inferior performance.

% 0 and
%0.01
.

\begin{figure}[t]
	\centering
	\includegraphics[width=0.5\linewidth]{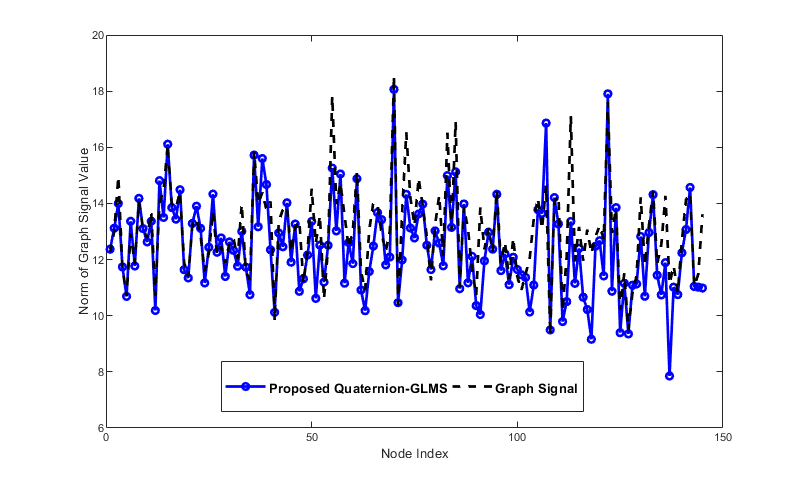}
	\caption{Reconstructed norm of the quaternion graph signal.}
		%		Reconstructed quaternion graph weather dataset signal. (a)-(d) show the reconstructed individual components of the quaternion signal, and (e) presents the reconstructed norm of the quaternion graph signal.
	\label{figweather}
\end{figure}

\textbf{Practical gains.}
	 We further validated QGLMS on a UK weather network with $N{=}145$ stations and $E{=}542$ edges; each node stores temperature, humidity, wind, and pressure as a quaternion sample.
		With $S{=}70$ randomly revealed nodes and a support size $|\mathcal F|{=}45$, QGLMS reconstructed a high-fidelity sparse band-limited signal. Figure \ref{figweather} shows the recovered sparse signal, demonstrating that QGLMS effectively captures the smooth spatial variations of multidimensional data.
\section{Conclusion}
In this paper, we introduced the QGLMS algorithm for adaptive graph signal recovery in the quaternion domain. By extending conventional LMS techniques to quaternion-valued signals on graphs, QGLMS effectively captures interdependencies between quaternion components, leading to improved reconstruction accuracy. Theoretical analysis provided the adaptation formula and convergence conditions, while simulations demonstrated the superiority of QGLMS over applying 4 separated RLMS filters to each quaternion component. QGLMS achieved lower NMSE and better signal recovery by leveraging the full quaternion structure.

%Overall, QGLMS offers a powerful framework for quaternion-valued graph signal processing with potential applications in multidimensional signal analysis. Future work may explore dynamic graph extensions and adaptive learning rate optimizations.
% or


\appendices
\section{Mean convergence }
Let $ \tilde{\mathbf{x}}[n] = \mathbf{x}[n] - \mathbf{x}^o $ be the error vector at
time $ n $. Subtracting $ \mathbf{x}^o$ from the left and right hand sides of \eqref{LMSFinal}, we have
\begin{align}\label{xtilde}
&\tilde{\mathbf{x}}[n+1] =
(\mathbf{I}  - 4\mu \mathbf{B}\mathbf{D}\mathbf{B})\tilde{\mathbf{x}}[n]-4\mu \mathbf{B}\mathbf{D}\mathbf{B}\tilde{\mathbf{x}}_0[n]  \nonumber \\
&+ 4\mu \mathbf{B}\mathbf{D}\mathbf{v}[n] + 4\mu  \mathbf{B}\mathbf{D}\mathbf{v}_0[n],
\end{align}
where we use $ \mathbf{B}\tilde{\mathbf{x}}= \tilde{\mathbf{x}} $. Applying QGFT  to the \eqref{xtilde} and $ \mathbf{B} = \mathbf{U}\mathbf{\Sigma}\mathbf{U}^H $, we have
\begin{align}\label{stilde}
&\tilde{\mathbf{s}}[n+1] =
(\mathbf{I}  - 4\mu \mathbf{\Sigma}\mathbf{U}^H\mathbf{D}\mathbf{U}\mathbf{\Sigma})\tilde{\mathbf{s}}[n]-4\mu \mathbf{\Sigma}\mathbf{U}^H\mathbf{D}\mathbf{U}\mathbf{\Sigma}\tilde{\mathbf{s}}_0[n]  \nonumber \\
&+ 4\mu\mathbf{\Sigma}\mathbf{U}^H\mathbf{D}\mathbf{v}[n] + 4\mu \mathbf{\Sigma}\mathbf{U}^H\mathbf{D}\mathbf{v}_0[n].
\end{align}

%So according to graph and letting $ \mathbf{U}_{\mathcal{F}} \in \mathbb{R}^{N \times |\mathcal{F}|} $, we have
%\begin{align}\label{shat}
%&\hat{\mathbf{s}}[n+1] =
%(\mathbf{I}  - 4\mu \mathbf{U}_{\mathcal{F}}^H\mathbf{D}\mathbf{U}_{\mathcal{F}})\hat{\mathbf{s}}[n]-4\mu \mathbf{U}_{\mathcal{F}}^H\mathbf{D}\mathbf{U}_{\mathcal{F}} \hat{\mathbf{s}}_0[n]  \nonumber \\
%&+ 4\mu \mathbf{U}_{\mathcal{F}}^H\mathbf{D}\mathbf{v}[n] + 4\mu \mathbf{U}_{\mathcal{F}}^H\mathbf{D}\mathbf{v}_0[n].
%\end{align}
%We can write the following equation for real and imaginary part of $\hat{\mathbf{s}}[n]$ as
%\begin{align}
% &\mathcal{R}\{\hat{\mathbf{s}}[n+1]\} =  \hat{\mathbf{s}}_0[n+1] = (\mathbf{I}  - 4\mu \mathbf{U}_{\mathcal{F}}^{H}\mathbf{D}\mathbf{U}_{\mathcal{F}})\hat{\mathbf{s}}_0[n] \nonumber \\
% &-4\mu \mathbf{U}_{\mathcal{F}}^H\mathbf{D}\mathbf{U}_{\mathcal{F}} \hat{\mathbf{s}}_0[n]+ 4\mu\mathbf{U}_{\mathcal{F}}^H\mathbf{D}\mathbf{v}_0[n] + 4\mu \mathbf{U}_{\mathcal{F}}^H\mathbf{D}\mathbf{v}_0[n] \nonumber \\
%& = (\mathbf{I}  - 8\mu \mathbf{U}_{\mathcal{F}}^H\mathbf{D}\mathbf{U}_{\mathcal{F}})\hat{\mathbf{s}}_0[n] +
% 8 \mu \mathbf{U}_{\mathcal{F}}^H\mathbf{D} \mathbf{v}_0[n]\nonumber \\
%&\mathcal{IM}\{\hat{\mathbf{s}}[n+1]\} =  (\mathbf{I}  - 4\mu \mathbf{U}_{\mathcal{F}}^H\mathbf{D}\mathbf{U}_{\mathcal{F}})\mathcal{IM}\{\hat{\mathbf{s}}[n]\} +‌ \nonumber \\
%& 4\mu \mathbf{U}_{\mathcal{F}}^H\mathbf{D}\mathcal{IM}\{\mathbf{v}[n]\}.
%\end{align}

With some calculations, the expected value of $ \hat{\mathbf{s}}[n] $ under the assumption of zero-mean  $ \mathbf{v}[n] $ is equal to
\begin{align}
&E\{\hat{\mathbf{s}}[n+1]\} =
(\mathbf{I}  - 4\mu \mathbf{U}_{\mathcal{F}}^H\mathbf{D}\mathbf{U}_{\mathcal{F}})E\{\hat{\mathbf{s}}[n]\}\nonumber \\
&-4\mu  \mathbf{U}_{\mathcal{F}}^H\mathbf{D} \mathbf{U}_{\mathcal{F}}E\{\hat{\mathbf{s}}_0[n]\}  \nonumber \\
&+ 4\mu \mathbf{U}_{\mathcal{F}}^H\mathbf{D} E\{\mathbf{v}[n]\} + 4\mu \mathbf{U}_{\mathcal{F}}^H\mathbf{D}E\{\mathbf{v}_0[n]\}
%\nonumber \\
%&E\{\hat{\mathbf{s}}_0[n+1] \}= (\mathbf{I}  - 8\mu  \mathbf{U}_{\mathcal{F}}^H\mathbf{D} \mathbf{U}_{\mathcal{F}})E\{\hat{\mathbf{s}}_0[n] \}\nonumber \\
%&E\{ \mathcal{IM}\{\hat{\mathbf{s}}[n+1]\}\} = (\mathbf{I}  - 4\mu  \mathbf{U}_{\mathcal{F}}^H\mathbf{D} \mathbf{U}_{\mathcal{F}})E\{\mathcal{IM}\{\hat{\mathbf{s}}[n]\}\}.
\end{align}
The above expression for real and imaginary parts leads to
\begin{align}
&\mathbb{E}\{\hat{\mathbf{s}}_0[n+1] \}= (\mathbf{I}  - 8\mu  \mathbf{U}_{\mathcal{F}}^H\mathbf{D} \mathbf{U}_{\mathcal{F}})^{n+1}\mathbb{E}\{\hat{\mathbf{s}}_0[0] \}\nonumber \\
&\mathbb{E}\{ \mathcal{IM}\{\hat{\mathbf{s}}[n+1]\}\} = (\mathbf{I}  - 4\mu  \mathbf{U}_{\mathcal{F}}^H\mathbf{D} \mathbf{U}_{\mathcal{F}})^{n+1}\mathbb{E}\{\mathcal{IM}\{\hat{\mathbf{s}}[0]\}\}.
\end{align}
This equations show that in order to guarantee convergence of the coefficients in the mean, the $ \mu $  must be chosen in the range indicated in (\ref{eq: mucond}).

%\begin{align}\label{23ref}
%\begin{cases}
%0 \le \mu \le \frac{1}{4 \lambda_{max}( \mathbf{U}_{\mathcal{F}}^H\mathbf{D} \mathbf{U}_{\mathcal{F}})},
%\\
%%\nonumber \\
%%\nonumber \\
%~~~~~~~~~~&  \Rightarrow  0 \le \mu \le \frac{1}{4 \lambda_{max}( \mathbf{U}_{\mathcal{F}}^H\mathbf{D} \mathbf{U}_{\mathcal{F}})}. \\
%% \nonumber \\
%0 \le \mu \le \frac{1}{2 \lambda_{max}( \mathbf{U}_{\mathcal{F}}^H\mathbf{D} \mathbf{U}_{\mathcal{F}})},
%\end{cases}
%\end{align}
\section{MSE convergence}
For MSE, we consider $ \|\hat{\mathbf{s}}[n]\|_{\mathbf{\Phi}}^2 = \hat{\mathbf{s}}[n]^H \mathbf{\Phi}\hat{\mathbf{s}}[n] $ for arbitrary Hermitian nonnegative-definite matrix $ \mathbf{\Phi} \in \mathbb{R}^{|\mathcal{F}| \times |\mathcal{F}x|} $ and then with some calculations with expandinbg $\mathbb{E}\{\|\hat{\mathbf{s}}[n]\|_{\mathbf{\Phi}}^2\}$, we will have
\begin{align}\label{18ref}
&\mathbb{E}\{\|\hat{\mathbf{s}}_0[n+1]\|_{\mathbf{\Phi}}^2\}= \mathbb{E}\{\mathbf{s}_0[n]\|_{\mathbf{\Phi}'}^2\}
+64 \mu^2 {\rm Tr}\{\mathbf{\Phi}(\mathbf{U}_{\mathcal{F}}^H\mathbf{D})\mathbf{C}_{\mathbf{v}_0}(\mathbf{D}\mathbf{U}_{\mathcal{F}})\}\nonumber \\
%&= \mathbb{E}\{\hat{\mathbf{s}}_0[n]^H(\mathbf{I}  - 8\mu  \mathbf{U}_{\mathcal{F}}^H\mathbf{D} %\mathbf{U}_{\mathcal{F}})\mathbf{\Phi}(\mathbf{I}  - 8\mu  \mathbf{U}_{\mathcal{F}}^H\mathbf{D} %\mathbf{U}_{\mathcal{F}})\hat{\mathbf{s}}_0[n]\} \nonumber \\
%&+64\mu^2 \mathbb{E}\{\mathbf{v}_0[n]^H(\mathbf{D}\mathbf{U}_{\mathcal{F}})\mathbf{\Phi}(\mathbf{U}_{\mathcal{F}}^H\mathbf{D})\mathbf{v}_0[n]\} \nonumber \\
%& \mathbb{E}\{\mathbf{s}_0[n]\|_{\mathbf{\Phi}'}^2\}
%+64 \mu^2 {\rm Tr}\{\mathbf{\Phi}(\mathbf{U}_{\mathcal{F}}^H\mathbf{D})\mathbf{C}_{\mathbf{v}_0}(\mathbf{D}\mathbf{U}_{\mathcal{F}})\}
% \nonumber \\
& \mathbb{E}\{\|\mathcal{IM}\{\hat{\mathbf{s}}[n+1]\}\|_{\mathbf{\Phi}}^2\} = \nonumber \\
%&=  \mathbb{E}\{\mathcal{IM}\{\hat{\mathbf{s}}[n]\}^H (\mathbf{I}  - 4\mu %\mathbf{U}_{\mathcal{F}}^H\mathbf{D}\mathbf{U}_{\mathcal{F}})\mathbf{\Phi}\nonumber \\
%&~~~~~~~~~~~~~~~~~(\mathbf{I}  - 4\mu \mathbf{U}_{\mathcal{F}}^H\mathbf{D}\mathbf{U}_{\mathcal{F}})\mathcal{IM}\{\hat{\mathbf{s}}[n]\}\}
%\nonumber \\
%&+ 16 \mu^2 %\mathbb{E}\{\mathcal{IM}\{\mathbf{v}[n]\}^H(\mathbf{D}\mathbf{U}_{\mathcal{F}})\mathbf{\Phi}(\mathbf{U}_{\mathcal{F}}^H\mathbf{D})\mathcal{IM}\{\mathbf{v}[n]\}\}
%\nonumber \\
&  \mathbb{E}\{\|\mathcal{IM}\{\hat{\mathbf{s}}[n]\}\|_{\mathbf{\Phi}''}^2\} + 16\mu^2 {\rm Tr}\{\mathbf{\Phi}(\mathbf{U}_{\mathcal{F}}^H\mathbf{D})\mathbf{C}_{\mathcal{IM}\{\mathbf{v}\}}(\mathbf{D}\mathbf{U}_{\mathcal{F}})\},
\end{align}
where $ \mathbf{\Phi}'=(\mathbf{I}  - 8\mu  \mathbf{U}_{\mathcal{F}}^H\mathbf{D} \mathbf{U}_{\mathcal{F}})\mathbf{\Phi}(\mathbf{I}  - 8\mu  \mathbf{U}_{\mathcal{F}}^H\mathbf{D} \mathbf{U}_{\mathcal{F}}) $ and $ \mathbf{\Phi}'' =  (\mathbf{I}  - 4\mu \mathbf{U}_{\mathcal{F}}^H\mathbf{D}\mathbf{U}_{\mathcal{F}})\mathbf{\Phi}(\mathbf{I}  - 4\mu \mathbf{U}_{\mathcal{F}}^H\mathbf{D}\mathbf{U}_{\mathcal{F}})$.
If we consider $ \mathbf{\phi} = {\rm vec}(\mathbf{\Phi}) $, $ \mathbf{\phi}' = {\rm vec}(\mathbf{\Phi}') $ and $ \mathbf{\phi}'' = {\rm vec}(\mathbf{\Phi}'') $
where the notation ${\rm vec} $
stacks the columns of matrix on top of each other and $ {\rm vec^{-1}()} $ is
the inverse operation. We will use interchangeably the notation $ \|\hat{\mathbf{s}}\|^2_{\mathbf{\Phi}} $ and $ \|\hat{\mathbf{s}}\|^2_{\phi} $ to denote the same quantity $ \hat{\mathbf{s}}^H\mathbf{\Phi}  \hat{\mathbf{s}} $. Exploiting
the Kronecker product property, we have $ {\rm vec}(\mathbf{X}\mathbf{\Phi}\mathbf{Y})  = (\mathbf{Y}^H \otimes \mathbf{X}){\rm vec}(\mathbf{\phi}) $ and the Trace property, we have $ {\rm Tr}(\mathbf{\Phi}\mathbf{X}) = {\rm vec}(\mathbf{X}^H)^T {\rm vec \mathbf{\Phi}},$ so from \eqref{18ref}, we have
\begin{align}\label{20ref}
&\mathbb{E}\{\|\hat{\mathbf{s}}_0[n+1]\|_{\mathbf{\phi}}^2\} =  \mathbb{E}\{\|\hat{\mathbf{s}}_0[n]\|_{\mathbf{Q}\mathbf{\phi}}^2\} + 64 \mu^2 {\rm vec}(\mathbf{G})^T\mathbf{\phi}\nonumber \\
&\mathbb{E}\{\|\mathcal{IM}\{\hat{\mathbf{s}}[n+1]\}\|_{\mathbf{\phi}}^2\} = \nonumber \\
& \mathbb{E}\{\|\mathcal{IM}\{\hat{\mathbf{s}}[n]\}\|_{\mathbf{Q}'\mathbf{\phi}}^2\} +16\mu^2 {\rm vec}(\mathbf{G'})^T\mathbf{\phi},
\end{align}
where $\mathbf{G} = (\mathbf{U}_{\mathcal{F}}^H\mathbf{D})\mathbf{C}_{\mathbf{v}_0}(\mathbf{D}\mathbf{U}_{\mathcal{F}}), \mathbf{G'} =  (\mathbf{U}_{\mathcal{F}}^H\mathbf{D})\mathbf{C}_{\mathcal{IM}\{\mathbf{v}\}}(\mathbf{D}\mathbf{U}_{\mathcal{F}})$, $\mathbf{Q} = (\mathbf{I}  - 8\mu  \mathbf{U}_{\mathcal{F}}^H\mathbf{D} \mathbf{U}_{\mathcal{F}}) \otimes (\mathbf{I}  - 8\mu  \mathbf{U}_{\mathcal{F}}^H\mathbf{D} \mathbf{U}_{\mathcal{F}})$, and $\mathbf{Q'} = (\mathbf{I}  - 4\mu \mathbf{U}_{\mathcal{F}}^H\mathbf{D}\mathbf{U}_{\mathcal{F}})\otimes (\mathbf{I}  - 4\mu \mathbf{U}_{\mathcal{F}}^H\mathbf{D}\mathbf{U}_{\mathcal{F}})$. Recursion \eqref{20ref} can be equivalently
recast as \eqref{25ref} with letting  $ \mathbf{r} = {\rm vec}(\mathbf{G}) $ and $ \mathbf{r}'={\rm vec}(\mathbf{G}') $. We have
%\begin{align} \label{22ref}
%& \mathbf{G} = (\mathbf{U}_{\mathcal{F}}^H\mathbf{D})\mathbf{C}_{\mathbf{v}_0}(\mathbf{D}\mathbf{U}_{\mathcal{F}}), \mathbf{G'} =  (\mathbf{U}_{\mathcal{F}}^H\mathbf{D})\mathbf{C}_{\mathcal{IM}\{\mathbf{v}\}}(\mathbf{D}\mathbf{U}_{\mathcal{F}}) \nonumber \\
%&\mathbf{Q} = (\mathbf{I}  - 8\mu  \mathbf{U}_{\mathcal{F}}^H\mathbf{D} \mathbf{U}_{\mathcal{F}}) \otimes (\mathbf{I}  - 8\mu  \mathbf{U}_{\mathcal{F}}^H\mathbf{D} \mathbf{U}_{\mathcal{F}}) \nonumber \\
%& \mathbf{Q'} = (\mathbf{I}  - 4\mu \mathbf{U}_{\mathcal{F}}^H\mathbf{D}\mathbf{U}_{\mathcal{F}})\otimes (\mathbf{I}  - 4\mu \mathbf{U}_{\mathcal{F}}^H\mathbf{D}\mathbf{U}_{\mathcal{F}}).
%%& \mathbf{G'} =  (\mathbf{U}_{\mathcal{F}}^H\mathbf{D})\mathbf{C}_{\mathcal{IM}\{\mathbf{v}\}}(\mathbf{D}\mathbf{U}_{\mathcal{F}})
%\end{align}

%\begin{align}
%\begin{cases}
%0 \le \mu \le \frac{1}{4 \lambda_{max}( \mathbf{U}_{\mathcal{F}}^H\mathbf{D} \mathbf{U}_{\mathcal{F}})}
%\\
%%\nonumber \\
%%\nonumber \\
%~~~~~~~~~~&  \Rightarrow  0 \le \mu \le \frac{1}{4 \lambda_{max}( \mathbf{U}_{\mathcal{F}}^H\mathbf{D} \mathbf{U}_{\mathcal{F}})} \\
%% \nonumber \\
%0 \le \mu \le \frac{1}{2 \lambda_{max}( \mathbf{U}_{\mathcal{F}}^H\mathbf{D} \mathbf{U}_{\mathcal{F}})}
%\end{cases}
%\end{align}
%\subsection* {Proof:}
\begin{align}\label{25ref}
&\mathbb{E}\{\|\hat{\mathbf{s}}_0[n]\|_{\mathbf{\phi}}^2\} =  \mathbb{E}\{\|\hat{\mathbf{s}}_0[0]\|_{\mathbf{Q}^n\mathbf{\phi}}^2\} + 64 \mu^2 \mathbf{r}^T\Sigma_{l=0}^{n-1}\mathbf{Q}^l\mathbf{\phi}\nonumber \\
&\mathbb{E}\{\|\mathcal{IM}\{\hat{\mathbf{s}}[n]\}\|_{\mathbf{\phi}}^2\} =  \\ \nonumber  &\mathbb{E}\{\|\mathcal{IM}\{\hat{\mathbf{s}}[0]\}\|_{\mathbf{Q}^{'n}\mathbf{\phi}}^2\} +16\mu^2\mathbf{r}^{'T}\Sigma_{l=0}^{n-1}\mathbf{Q}^{'l}\mathbf{\phi},
\end{align}
where $ \mathbb{E}\{\|\hat{\mathbf{s}}_0[0]\|\} $  and $\mathbb{E}\{\|\mathcal{IM}\{\hat{\mathbf{s}}[0]\}\|\} $ denotes the initial condition for real and imaginary parts of $ \mathbf{s} $. We first note that if $ \mathbf{Q} $ and  $ \mathbf{Q}' $ are  stable $ \mathbf{Q}^{n \rightarrow \infty} \rightarrow 0  $ and $ \mathbf{Q}^{' n \rightarrow \infty} \rightarrow 0 $. Now, with some calculations and discussions omitted here for brevity, condition guaranteing the stability of $\mathbf{Q} $ and $ \mathbf{Q}'$ are that $ \|(\mathbf{I}  - 8\mu \mathbf{U}_{\mathcal{F}}^H\mathbf{D}\mathbf{U}_{\mathcal{F}})\| \le 1   $ and $ \|(\mathbf{I}  - 4\mu \mathbf{U}_{\mathcal{F}}^H\mathbf{D}\mathbf{U}_{\mathcal{F}})\| \le 1 $, which hold true for any step-sizes satisfying \eqref{eq: mucond}.

\ifCLASSOPTIONcaptionsoff
  \newpage
\fi

\end{document}